*Article*

# The Riemannian Means Field Classifier for EEG-Based BCI Data

Anton Andreev [1], Gregoire Cattan [2] and Marco Congedo [1,*]

[1] GIPSA-lab, Université Grenoble Alpes, CNRS, Grenoble INP, 38000 Grenoble, France; andreev.anton@gipsa-lab.grenoble-inp.fr
[2] IBM, Data and AI, 30-150 Krakow, Poland; gregoire.cattan@ibm.com
* Correspondence: marco.congedo@gipsa-lab.grenoble-inp.fr

**Abstract:** A substantial amount of research has demonstrated the robustness and accuracy of the Riemannian minimum distance to mean (MDM) classifier for all kinds of EEG-based brain–computer interfaces (BCIs). This classifier is simple, fully deterministic, robust to noise, computationally efficient, and prone to transfer learning. Its training is very simple, requiring just the computation of a geometric mean of a symmetric positive-definite (SPD) matrix per class. We propose an improvement of the MDM involving a number of power means of SPD matrices instead of the sole geometric mean. By the analysis of 20 public databases, 10 for the motor-imagery BCI paradigm and 10 for the P300 BCI paradigm, comprising 587 individuals in total, we show that the proposed classifier clearly outperforms the MDM, approaching the state-of-the art in terms of performance while retaining the simplicity and the deterministic behavior. In order to promote reproducible research, our code will be released as open source.

## 1. Introduction

Among the various classifiers explored in the field of brain–computer interfaces (BCIs), those leveraging Riemannian geometry have consistently demonstrated high performance, as highlighted in a recent extensive benchmark study [1]. These algorithms operate efficiently and require relatively less data for training as compared to competitive alternatives, such as deep learning models [2]. Riemannian geometry, an area of differential geometry, examines smooth manifolds—curved spaces with distinctive geometric structures. Within these spaces, fundamental concepts such as angles, geodesic paths, distances, and centers of mass enable a geometric interpretation of mathematical operators, making them more intuitive to analyze [3]. In BCI research, the manifold of symmetric positive-definite (SPD) matrices [4] has proven particularly valuable, as multivariate electroencephalography (EEG) data collected over finite time windows can be effectively mapped as points on this manifold by computing their covariance matrices [5–13]. This methodology has led to the development of classifiers with significant advantages over conventional approaches [1]. For a detailed introduction to the SPD Riemannian manifold, readers are referred to [4], while an overview of its applications in BCIs is provided in [8] and [9].

Riemannian classifiers have demonstrated high accuracy, robustness to noise, and generalizability, outperforming competing methods [1] and winning multiple international BCI machine learning competitions [8]. One classifier of particular importance is the Riemannian minimum distance to mean (MDM), which serves as the foundational model for all Riemannian classifiers acting directly on the manifold [5]. The MDM classifier is highly adaptable to multi-user scenarios [10], effective for individuals with clinical conditions [11], and well-suited for transfer learning and adaptive strategies, as exemplified by the calibration-free P300-based BCI video game Brain Invaders [12,13]. Its versatility stems from several key characteristics, including simplicity, computational efficiency, and complete determinism, meaning it is entirely free from hyperparameters. Because the MDM classifier's performance is not influenced by nuisance parameters, it serves as an ideal benchmark for evaluating preprocessing and processing techniques. In fact, when evaluating these techniques within classification pipelines, the result must not depend on the choice of hyperparameters and the classifier must be as general as possible. Although it is not inherently the most accurate among Riemannian classifiers, its performance is significantly enhanced when covariance matrices are estimated using high-dimensional phase-space embedding [7]. However, this enhancement comes with a large hyperparameters search, disabling the classifier's desirable properties. Thus, in order to improve the MDM while keeping determinism, we follow here a different path.

The effectiveness of the MDM classifier primarily stems from its use of a metric that is well-suited to the structure of the symmetric positive-definite (SPD) manifold. This metric governs both the way distances between two points are measured and how the center of mass is defined for a set of points, since the center of mass is characterized as the point on the manifold that minimizes the overall dispersion of the point cloud relative to itself [8,14]. Up to now, the hyperbolic (geometric) distance and the geometric mean, both of which are induced by the Fisher–Rao (affine-invariant) metric, have been the preferred choices because of the desirable invariance properties they exhibit. As discussed in [8], these represent the extensions of the classical hyperbolic distance and geometric mean for scalars to the domain of SPD matrices.

In [15], the authors introduced a one-parameter family of means (parameterized by $h$) that extends the notion of power means for scalars to symmetric positive-definite (SPD) matrices. Similar to scalar power means, these SPD matrix power means [15] (see also [16–18]) provide a continuum of means interpolating between the harmonic mean ($h = -1$) and the arithmetic mean ($h = 1$), with the geometric mean emerging as the limit case when $h$ approaches zero from either side. This flexible generalization proves valuable in the context of brain–computer interfaces (BCIs), as noted in [19], where electroencephalogram (EEG) recordings are typically contaminated with various noise sources. By adjusting $h$, it is possible to identify a mean that is best suited to the specific signal-to-noise ratio of the data. In [19], the authors evaluated the performance of the minimum distance to mean (MDM) classifier using 13 different power means with $h = \{\pm 1, \pm 0.8, \pm 0.6, \pm 0.4, \pm 0.2, \pm 0.1, 0\}$ (refer to Figure 7 in [19]). The standard MDM classifier employs the geometric mean, corresponding to the limit case $h \to 0$. Their analysis revealed that although values of $h$ close to zero tended to yield high classification accuracy, the geometric mean itself was the optimal choice for only three out of thirty-eight subjects tested. Instead, the ideal $h$ varied significantly among individuals, suggesting that while the geometric mean is a principled guess, it is not universally optimal. Notably, there was a clear positive correlation between the highest achieved classification accuracy and the corresponding optimal $h$: better-performing classifications tended to align with higher optimal $h$ values, indirectly indicating that the optimal mean depends on the underlying signal-to-noise characteristics. As pointed out in [19], determining the best $h$ for a given subject and session requires supervised learning, thereby reintroducing challenges such as overfitting, a lack of transfer learning, and the need for hyperparameter tuning. However, this insight motivated the introduction of the minimum distance to means field (MDMF) classifier [20], which overcomes these limitations by considering a range of power means within the interval $h \in [-1, 1]$. Like the traditional MDM, MDMF remains free of hyperparameters but operates using multiple means instead of solely relying on the geometric mean (see Figure 1). In the present article, we go a step further and we propose the means field (MF) classifier. Unlike MDMF, which uses the minimum distance to several means, MF submits the squared distances to all sampled power means—referred to as a "means field"—to a classifier that is trained jointly with these means (as illustrated in Figure 1). Provided that this classifier does not require hyperparameter tuning, the MF approach also remains hyperparameter-free, except for the initial selection of the means field. Nevertheless, as long as the means field offers a representative set of power means sampled along the interval [−1, 1], this choice has minimal impact on overall performance.

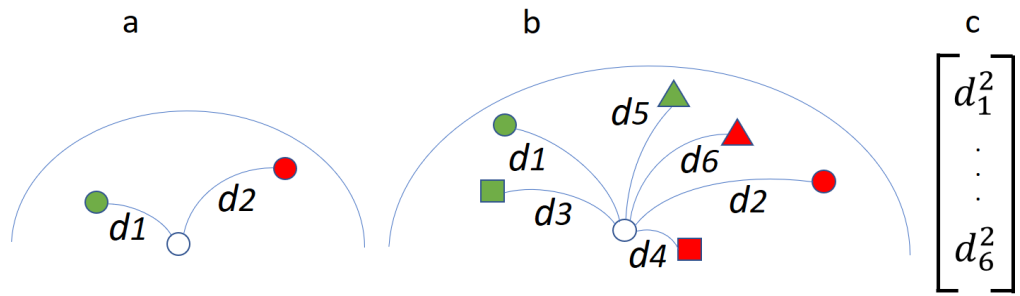

**Figure 1.** Schematic representation of the operation of the MDM, MDMF, and MF algorithms on the manifold of positive-definite matrices for a two-class problem. (**a**) MDM algorithm; the colored disks represent the geometric means for the two classes (red and green) as learned during the training phase. The circle is an unlabeled trial to be classified. The curves joining the disks and the circle represent the distances between them along geodesics. In this example, the trial would be classified as belonging to the green class as distance $d_1$ is the shortest. (**b**) Similar to (**a**), but employing a mean field composed of three power means per class. The MDMF [20] classifies the trial as belonging to the red class, as now distance $d_4$ is the shortest. (**c**) The MF submits the feature vector comprising the square of the distances $d_1$ to $d_6$ to a classifier that has been trained along with the power means. While the MDMF must classify using the minimum of all distances, the MF can learn any complex function thereof. For all three algorithms, the operation is the same for any number of classes and any number of power means per class.

The key contribution of this work is the introduction of the MF classifier, which is capable of learning complex distance patterns directly on the Riemannian manifold—patterns that similar previous methods, such as MDM and MDMF, fail to capture. Notably, the four best-performing algorithms reported in [1] are based on tangent space (TS) representations, in contrast to the MDM, MDMF and the MF introduced here, which all operate on the manifold. In addition, we propose several methodological improvements over [20], including a robust estimation of power means and an adaptive spatial filtering technique based on the well-established common spatial pattern (CSP) method. With these enhancements, and through the introduction of the MF classifier, we demonstrate that performance levels approaching those of current state-of-the-art TS-based methods can be achieved by Riemannian classifiers acting directly on the manifold.

Using the MOABB (Mother of All BCI Benchmark, v 1.1.1) [21,22] and the pyRiemann packages (v0.7) [23], we benchmark the MF against the MDMF, the MDM, and the logistic regression classifier with L2 penalty on the Riemannian tangent space, the latter being considered superior to Riemannian classifiers that act directly on the manifold. The benchmark is run on 20 databases covering two BCI paradigms, motor imagery (MI), and P300, for a total of 587 individuals. MOABB is an open-source framework for objectively assessing the performance of BCI classifiers on large amount of data. The use of MOABB ensures that exactly the same processing pipeline is applied to all databases of the same type and that both the cross-validation procedure and the Riemannian classifiers operate exactly in the same way for all databases, regardless the BCI type.

The rest of this article is organized as it follows. Section 2 describes the materials and methods, including our preprocessing and processing steps, our contributions, the data used for generating the results, and the statistical analysis we have planned. Section 3 presents our results for two types of BCI data: MI and P300. Section 4 contains the discussion, and Section 5 concludes the paper.

## 2. Materials and Methods

*2.1. Preprocessing, Processing, and Feature Extraction*

We use the default preprocessing of MOABB. The band-pass region is 8–32 Hz for the MI paradigm and 1–24 Hz for the P300 paradigm. Filtering is operated by means of a 4th-order Butterworth bandpass IIR filter. As done in [1], all available trials were used without any artifact rejection.

The first step in a Riemannian classifier is the estimation of a covariance matrix for each trial. This is standard processing in MOABB and has been detailed elsewhere [6,8]. We use the Oracle Approximating Shrinkage (OAS) estimator to ensure that the estimated covariance matrices are positive definite, a necessary condition for subsequent processing on the Riemannian manifold.

As compared to [20], we have improved the estimation of power means by implementing a robust algorithm, as detailed in Section 2.1.2. For MI data, we also propose an Adaptive Double Common Spatial Pattern (ADCSP) spatial filter, detailed in Section 2.1.1. In the results section, we provide a comparison of the ADCSP with the Common Spatial Filter CSP [24] with four filters per class. For P300 data, we employ the Xdawn spatial filter [24,25], which is the equivalent of the CSP (which applies to event-related (de)synchronization) for event-related potentials data. Four filters per class have been retained for Xdawn as well.

2.1.1. ADCSP

It is known that the MDM classifiers do not work well with a large number of electrodes, that is, with large covariance matrices [2]. Therefore, a spatial filter [6] is often employed as a first step in processing EEG-based BCI datasets in order to reduce the dimensionality and improve the class separability. Our ADCSP procedure for MI data is built using the CSP implementation available in pyRiemann. A CSP algorithm is applied in two stages, where the first allows a fast dimensionality reduction and the second, if needed, finds the optimal spatial filter to enhance class separability. The mean covariance matrix for each class is estimated using the Euclidean and geometric mean in the first and second stage, respectively. The means estimated this way are jointly diagonalized by means of a generalized eigenvalue-eigenvector decomposition [6] and Pham's approximate joint diagonalization algorithm [26] in the first and second stage, respectively. Stage 1 is entered if the dimension of the covariance matrices is ≥ 28 and reduces the dimension to 28. The second stage, always executed after the first, is entered if the dimension ≥ 10 and reduces the dimension to 10. The two-stage procedure of the ADCSP guarantees a maximum dimension of 10 for the covariance matrices to be submitted for classification.

2.1.2. Robust Power Mean Estimation (RPME)

The mean field for this study is defined as the set of 11 power means for each class with $h$ = {±1, ±0.75, ±0.5, ±0.25, ±0.1, 0}. Power means are obtained by the fixed point algorithm detailed in [19]. The pyRiemann implementation of this algorithm at the time of this research has been improved in several ways.

First, in order to enhance robustness with respect to outliers, the iterative strategy first proposed in [27] is adopted; the algorithm is run up to $I$ times, whereas after each time all outliers are removed from the set, until no more outliers are found or the number of iterations = $I$. In this study, $I$ has been fixed to four and outliers are defined as those trials in which the standardized geometric distance [6] from the mean exceeds a threshold $z_{th}$ = 2.5.

Second, since the power means are computed for several closely spaced values of $h$, the overall calculations can be hastened by computing the power means in increasing or decreasing order of $h$ and initializing the algorithm with the previous result. Remember that the power means for $h$ = (1, lim $h \to 0$, −1) are the usual Euclidean, geometric, and harmonic means, respectively.

2.1.3. Classification Pipelines

The MDM learns the geometric mean of each class using $k$-fold cross-validation. An unlabeled trial is assigned to the class of the mean associated with the minimum distance (Figure 1). The MDMF instead learns a mean field, i.e., a number of power means for each class with parameter $h$ sampling the [−1, 1] continuum. Again, an unlabeled trial is assigned to the class of the mean associated with the minimum distance (Figure 1). The MF estimates the power means as the MDMF does. However, while both the MDM and the MDMF rely on a single distance (the minimum distance) to classify, the MF classifies feature vectors formed by the squared distances of the training trials to all means (Figure 1). As a classifier for the MF, in this study, we use the linear discriminant analysis (LDA), which executes quickly and does not need hyperparameter tuning. To compute the distance between trials and the means composing the means field (Figure 1) we always adopt the usual affine-invariant metric [6].

These three algorithms (MDM, MDMF, and MF) are tested with and without the ADCSP. The logistic regression in Riemannian tangent space (TS + LR) [5] is also tested, as it is the state-of-the-art Riemannian classifier and, indeed, in [1] it was found to display excellent performance, typically significantly better than the MDM.

*2.2. Data*

MI and P300 are two widely studied paradigms in the BCI field, where MI relies on the mental imagination of movement to generate distinctive brain activity patterns, and P300 is based on detecting event-related potentials elicited by infrequent target stimuli in an oddball paradigm. Tables 1 and 2 report the main characteristics of the 20 databases we have used for testing. All databases are open-access and concern EEG-based BCI data. Ten databases are based on the MI BCI paradigm and 10 on the P300 BCI paradigm. Both tables present references in the first column that point to detailed information on the experimental conditions and procedures for each database. Additionally, MOABB provides a brief summary for each database (https://neurotechx.github.io/moabb/dataset_summary.html, accessed on 1 February 2025). For some databases, some of the subjects underwent more than one session, thus the actual number of sessions analyzed is superior to the total number of subjects. The number of electrodes used in the databases is highly variable for MI databases, ranging from three to 128. For P300 databases, it is rather homogeneous instead, ranging from 8 to 32. Since the number of available channels is low for P300 data, we did not implement an adaptive Xdawn procedure akin to the ADCSP.

Table 1. Main characteristics of motor imagery datasets.

| Name | Number of Channels | Number of Sessions | Number of Subjects |
|---|---|---|---|
| BNCI2014-001 [28] | 22 | 2 | 9 |
| BNCI2014-004 [29] | 3 | 5 | 9 |
| Cho 2017 [30] | 64 | 1 | 52 |
| Grosse Wentrup 2009 [31] | 128 | 1 | 10 |
| Lee2019 MI [32] | 62 | 2 | 54 |
| Physionet Motor Imagery [33] | 64 | 1 | 109 |
| Schirrmeister 2017 [34] | 128 | 1 | 14 |
| Shin 2017A [35] | 30 | 3 | 29 |
| Weibo 2014 [36] | 60 | 1 | 10 |
| Zhou 2016 [37] | 14 | 3 | 4 |

Table 2. Main characteristics of P300 datasets.

| Name | Number of Channels | Number of Sessions | Number of Subjects |
|---|---|---|---|
| BNCI2014-008 [38] | 8 | 1 | 8 |
| BNCI2014-009 [39] | 16 | 3 | 10 |
| BNCI2015-003 [40] | 8 | 1 | 10 |
| Brain Invaders 2012 [41] | 16 | 2 | 25 |
| Brain Invaders 2013a [42] | 16 | 1/8 | 24 |
| Brain Invaders 2014a [12] | 16 | 3 | 64 |
| Brain Invaders 2014b [43] | 32 | 3 | 38 |
| Brain Invaders 2015a [44] | 32 | 3 | 43 |
| Brain Invaders 2015b [45] | 32 | 1 | 44 |
| Cattan 2019 VR [46] | 16 | 2 | 21 |

### 2.3. Statistical Analysis

The different pipelines we considered were benchmarked using the statistical framework implemented in MOABB [22] using the Area Under the Receiving Operating Characteristics Curve (AUC-ROC) as a classification performance index estimated via a within-session stratified 5-fold cross-validation. Folds are kept identical for all pipelines. In a nutshell, within each database, the difference in central tendency of the AUC-ROC in two pipelines is tested by means of the following:

- A paired permutation one-sided test enumerating all possible data permutations if the number of subjects is <20, yielding in this case an exact test [47,48];
- The Wilcoxon one-sided signed-rank test, which basically is equivalent to a permutation test performed on the ranked data, if the number of subjects ≥ 20.

The *p*-values thus obtained for each database are combined across databases using the weighted Liptak combination function [48] with weights taken as the square root of the number of subjects in each database. This returns a single *p*-value for the global one-sided comparison of the central tendency of two pipelines. The effect sizes were also estimated following an approach similar to that used in meta-analysis studies. Specifically, the standardized mean difference (SMD) was computed within each database and subsequently aggregated across databases using a weighted arithmetic mean, employing the same weights as those applied in Liptak's *p*-value combination method.

We also report the execution "time" given in MOABB as an estimate of the time in seconds spent per fold on both training and classification. Note that the time heavily depends on CPU characteristics, including the number of logical processors available for multi-threading. Thus, execution time is given here just as a rough indication and no statistical analysis is performed on this variable. The benchmark was generated using a Dell laptop equipped with an Intel CPU i7-11850H @ 250 GHz (8 cores, 16 threads) and 32 GB of RAM.

## 3. Results

### 3.1. Motor Imagery

Table 3 presents the mean and standard deviation AUC-ROC score and mean time across all sessions and all databases for all pipelines. The proposed MF classifier with ADCSP with and without RPME gives the two best average AUCs, followed closely by the golden-standard TS + LR. The standard (i.e., without spatial filtering) MDM and MDMF pipelines feature the two worst performances, far behind. When it comes to execution time, the MDM and TS + LR with ADCSP and the MF with CSP are the fastest.

**Table 3.** Mean and StD AUC-ROC score and mean time per pipeline across motor imagery databases.

| Pipeline | Mean AUC-ROC | StD AUC-ROC | Mean Time |
|---|---|---|---|
| **ADCSP + MDM** | 0.685 | 0.187 | 0.15 |
| **ADCSP + MDMF** | 0.699 | 0.189 | 0.188 |
| **ADCSP + MF** | 0.765 | 0.18 | 0.186 |
| **ADCSP + MF_RPME** | 0.766 | 0.177 | 0.247 |
| **ADCSP + TS + LR** | 0.757 | 0.189 | 0.143 |
| **CSP + MF** | 0.728 | 0.187 | 0.149 |
| **MDM** | 0.655 | 0.185 | 0.251 |
| **MDMF** | 0.668 | 0.189 | 2.884 |
| **MF** | 0.756 | 0.175 | 2.58 |
| **TS + LR** | 0.763 | 0.188 | 0.264 |
| *Mean* | *0.724* | *0.185* | - |
| *StD* | *0.044* | *0.005* | - |

The detailed results per database for pipeline MDM, MDMF, MF, and TS + LR, all with ADCSP, are given in Table 4.

**Table 4.** Mean AUC-ROC score for each database for pipeline MDM, MDMF, MF, and TS + LR, all with ADCSP.

| Pipeline<br>Database | ADCSP + MDM | ADCSP + MDMF | ADCSP + MF | ADCSP + TS + LR |
|---|---|---|---|---|
| **BNCI2014-001** | 0.821 | 0.833 | 0.856 | 0.862 |
| **BNCI2014-004** | 0.777 | 0.783 | 0.796 | 0.801 |
| **Cho2017** | 0.672 | 0.689 | 0.737 | 0.745 |
| **GrosseW.2009** | 0.74 | 0.765 | 0.857 | 0.864 |
| **Lee2019-MI** | 0.728 | 0.747 | 0.822 | 0.826 |
| **PhysionetMI** | 0.592 | 0.599 | 0.682 | 0.66 |
| **Schirrm.2017** | 0.756 | 0.78 | 0.867 | 0.883 |
| **Shin2017A** | 0.64 | 0.655 | 0.722 | 0.692 |
| **Weibo2014** | 0.65 | 0.677 | 0.816 | 0.83 |
| **Zhou2016** | 0.906 | 0.908 | 0.931 | 0.941 |
| *Mean* | *0.728* | *0.744* | *0.809* | *0.81* |
| *StD* | *0.094* | *0.091* | *0.076* | *0.088* |

The SMD and *p*-values for the statistical tests contrasting selected pairs of pipelines within each database as well as the meta-effect combining them all across databases (see Section 2.3) are shown in Figures 2–4.

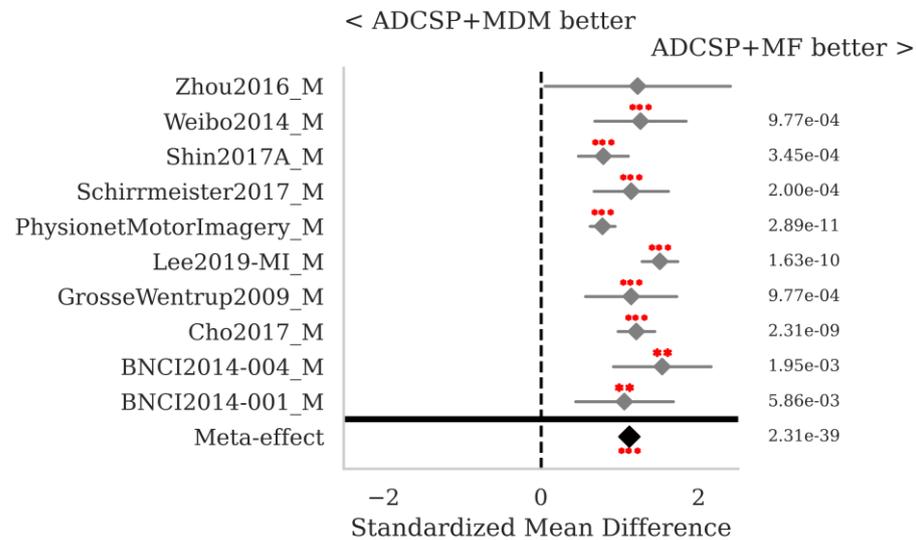

**Figure 2.** MI data: standardized mean differences (grey diamonds) with 95% confidence interval (grey horizontal lines) and *p*-values per database. The comparison concerns the MF with ADCSP and the MDM with ADCSP pipelines. A positive SMD value indicates that the algorithm on the right is better and the opposite if the value is negative. The closer the SMD is to zero, the lower the effect size. The *p*-values for the statistical test described in Section 2.3 are shown in the right column. If significant, one, two, or three red dots indicate a *p*-value < 0.05, <0.01, or <0.001, respectively. The meta-effect at the bottom of the figure combines the SMD and *p*-values across all databases.

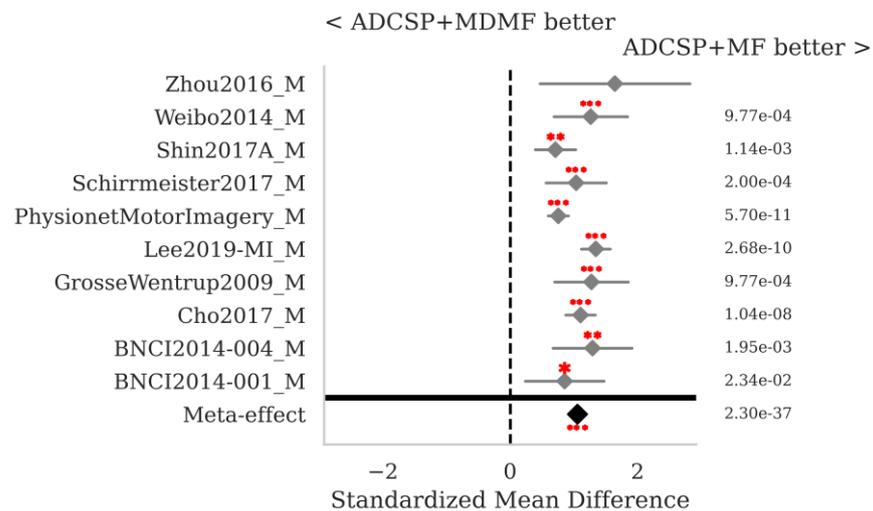

**Figure 3.** MI data: comparison of the MF and MDMF pipeline, both with ADCSP. See Figure 2 for the legend.

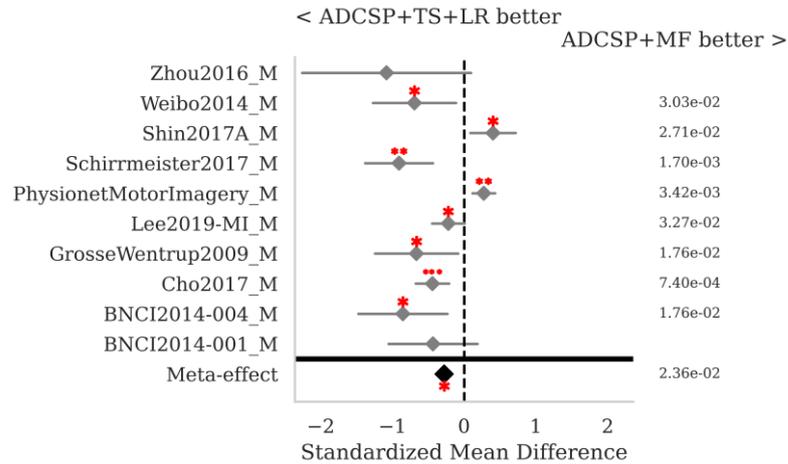

**Figure 4.** MI data: comparison of the MF with ADCSP and TS + LR with ACSTP pipeline. See Figure 2 for the legend.

The proposed MF classification pipeline with ADCSP significantly outperforms the MDM with ADCSP (Figure 2) and the MDMF with ADCSP (Figure 3) for all databases taken individually and overall, with meta-effect SMD = 1.122 ($p < 0.001$) and 1.057 ($p < 0.001$), respectively.

Comparing the MF pipeline to the golden-standard TS + LR, both applying the ADCSP, reveals that the latter performs significantly better for six out of ten databases, while the former performs significantly better for two out of ten databases; the SMD meta-effect is −0.28 ($p = 0.024$) in favor of TS + LR (Figure 4). Note that these comparisons are performed differently from those reported in Table 3; here, a $p$-value is obtained for each database and combined across databases, while in Table 3 the mean is computed across all sessions in all databases.

Table 1 shows that six out of ten MI databases comprise 60 or more electrodes. We have noticed that the MF is slow on these databases. It is well known that the execution time for MI data can be significantly reduced by using CSP filtering and in this article, we have proposed an improvement named ADCSP. Figure 5 shows the comparison of the MF pipeline with the standard CSP as implemented in pyRiemann and the proposed ADCSP filtering. The ADCSP performs better in six out of ten databases. The meta-effect is SMD = 0.453 ($p < 0.001$).

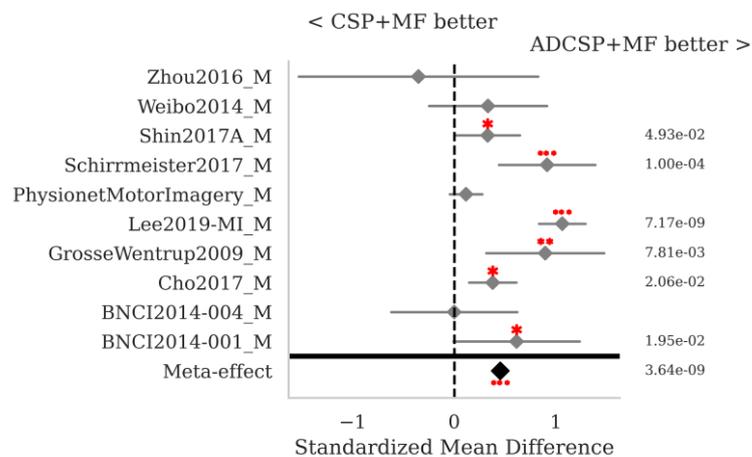

**Figure 5.** MI data: comparison of the MF pipeline, with ADCSP and the standard CSP. See Figure 2 for the legend. For BNCI2014-001, the results are identical, consequently resulting in an effect size of zero.

Figure 6 shows that MF with ADCSP and RPME is slightly better than MF with ADCSP only (SMD = −0.11, *p* < 0.01); thus, the robust power mean estimation does help in improving the robustness of the pipeline. However, this comes at the price of an increase in execution time (Table 3).

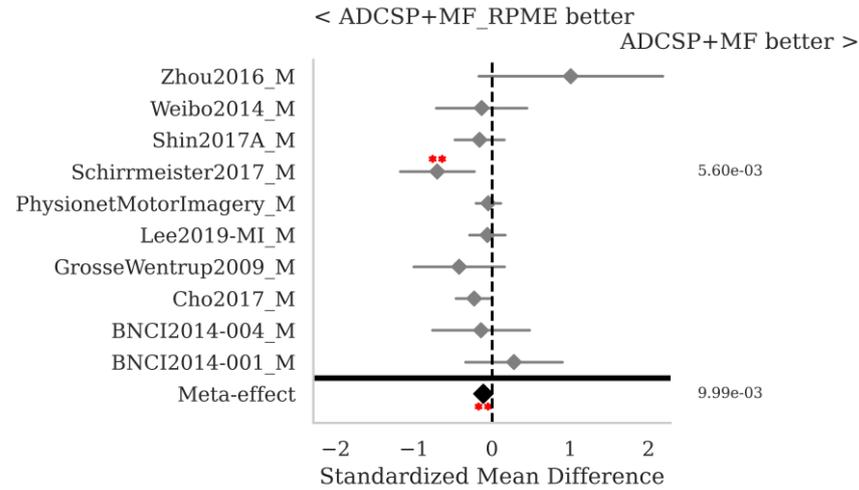

**Figure 6.** MI data: comparison of the MF with ADCSP, with and without RPME. See Figure 2 for the legend.

Table 3 shows that the ADCSP step contributes to the performance of the proposed pipeline, nonetheless, the MF performs only slightly lower without ADCSP; thus, the new classification scheme adopted by the MF algorithm appears to play the major role for its superiority. While improving the classification accuracy, the ADCSP also speeds up the execution time considerably.

*3.2. P300 ERPs*

Table 5 presents the mean and standard deviation AUC-ROC score and average time across all sessions and all databases for all pipelines. Let us remind that all P300 pipelines include Xdawn filtering, which is the counterpart of CSP for ERP data, and therefore these pipelines for P300 are all in all comparable to those employed for MI including CSP or ADCSP filtering. In contrast with what we have observed for MI data (Table 3), the mean AUC-ROC score is very close for all pipelines. The standard deviation is sensibly higher for MDM and MDMF as compared to the other pipelines. The golden-standard TS + LR displays the highest mean AUC-ROC, with the proposed MF with RPME a close second.

**Table 5.** Mean and StD AUC-ROC score and mean execution time per pipeline across P300 databases. All pipelines use Xdawn.

| Pipeline | MeanAUC ROC | StDAUC-ROC | Mean Time |
|---|---|---|---|
| **MDM** | 0.879 | 0.097 | 0.178 |
| **MDMF** | 0.879 | 0.095 | 1.064 |
| **MF** | 0.89 | 0.088 | 0.955 |
| **MF_RPME** | 0.893 | 0.086 | 2.823 |
| **TS + LR** | 0.898 | 0.084 | 0.181 |
| *Mean* | *0.888* | *0.09* | - |
| *StD* | *0.009* | *0.006* | - |

The detailed results per database for pipeline MDM, MDMF, MF, and TS + LR for all databases are given in Table 6.

**Table 6.** Mean AUC-ROC score for each database for pipeline MDM, MDMF, MF, and TS + LR, all with Xdawn.

| Pipeline Dataset | Xdawn + MDM | Xdawn + MDMF | Xdawn + MF | Xdawn + TS + LR |
|---|---|---|---|---|
| **BNCI2014-008** | 0.776 | 0.797 | 0.831 | 0.858 |
| **BNCI2014-009** | 0.92 | 0.926 | 0.931 | 0.93 |
| **BNCI2015-003** | 0.831 | 0.829 | 0.834 | 0.838 |
| **BrainInv.2012** | 0.882 | 0.883 | 0.901 | 0.907 |
| **BrainInv.2013a** | 0.91 | 0.914 | 0.922 | 0.922 |
| **BrainInv.2014a** | 0.809 | 0.807 | 0.834 | 0.857 |
| **BrainInv.2014b** | 0.916 | 0.916 | 0.914 | 0.913 |
| **BrainInv.2015a** | 0.926 | 0.923 | 0.925 | 0.927 |
| **BrainInv.2015b** | 0.835 | 0.834 | 0.839 | 0.843 |
| **Cattan2019-VR** | 0.899 | 0.901 | 0.913 | 0.913 |
| *Mean* | *0.87* | *0.873* | *0.884* | *0.891* |
| *StD* | *0.053* | *0.051* | *0.044* | *0.037* |

The SMD and *p*-values for the statistical tests contrasting selected pairs of pipelines within each database as well as the meta-effect combining them all across databases (see Section 2.3) are shown in Figures 7–10. The proposed MF classification pipeline significantly outperforms the standard MDM (Figure 7) and MDMF (Figure 8) algorithms for seven and eight out of the ten databases individually, respectively, and overall with a meta-effect SMD = 0.522 ($p < 0.001$) and SMD = 0.681 ($p < 0.001$), respectively.

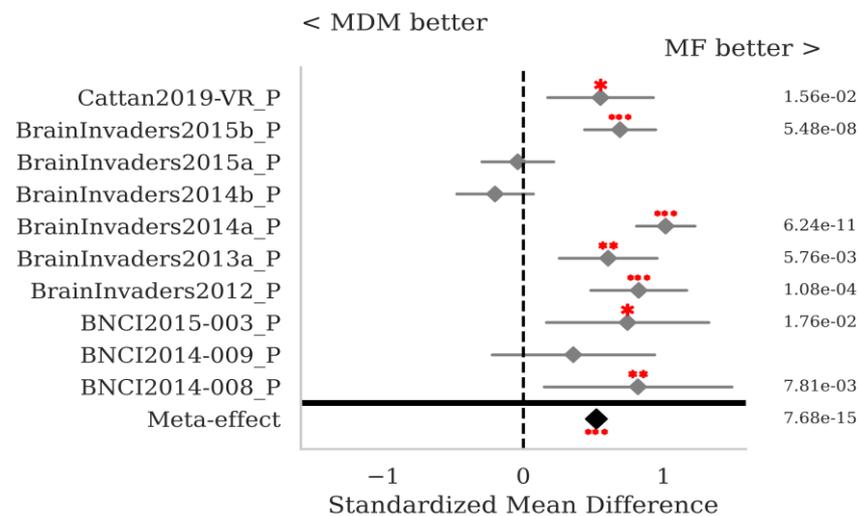

**Figure 7.** P300 data: comparison of the proposed MF to the MDM, both with Xdawn. See Figure 2 for the legend.

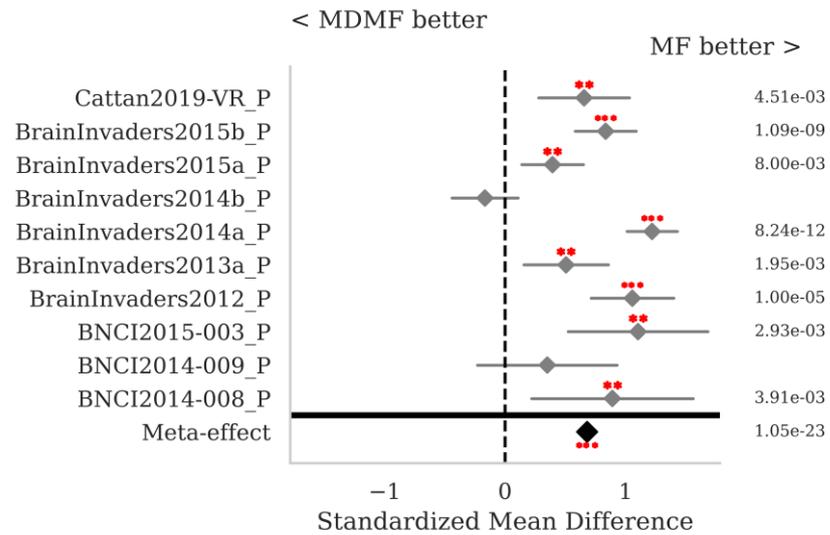

**Figure 8.** P300 data: comparison of the proposed MF to the MDMF pipelines, both with Xdawn. See Figure 2 for the legend.

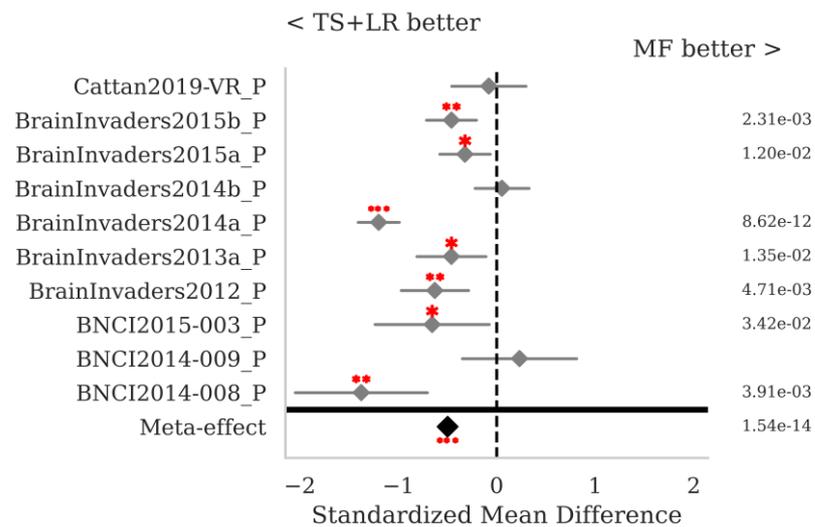

**Figure 9.** P300 data: comparison of the proposed MF to the golden-standard TS + LR pipelines, both with Xdawn. See Figure 2 for the legend.

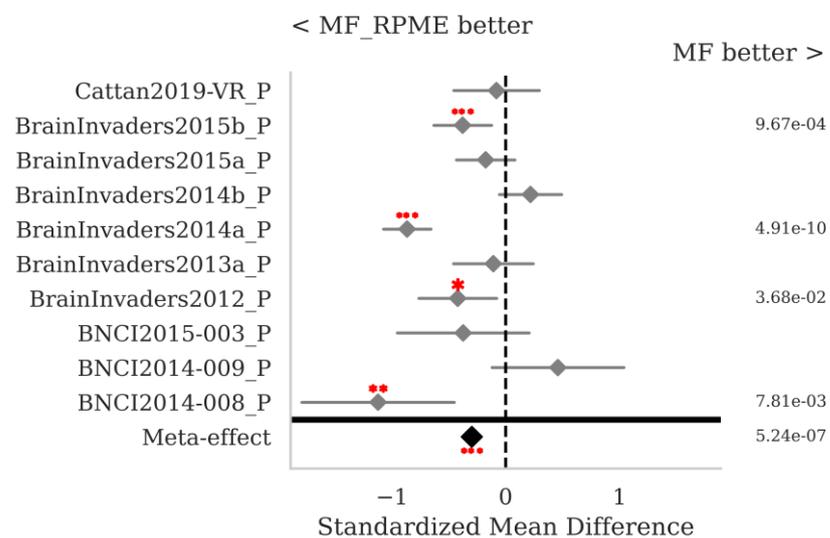

**Figure 10.** P300 data: comparison of the proposed MF pipeline with Xdawn, with and without RPME. See Figure 2 for the legend.

Comparing the MF pipeline to the gold standard TS + LR reveals that the latter performs significantly better for seven out of ten databases individually, and overall with a meta-effect SMD = −0.498 ($p$ < 0.001) (Figure 9). Note that these comparisons are performed differently from those reported in Table 5; here, a $p$-value is obtained for each database and combined across databases, while in Table 5 the mean is computed across all sessions in all databases.

Figure 10 shows that MF with RPME is, in the case of P300 data, clearly better than MF without RPME (SMD = −0.297, $p$ < 0.001), thus the robust power mean estimation clearly helps in this case, improving the robustness of the pipeline.

## 4. Discussion

In this article, we have proposed an improvement of the standard Riemannian MDM algorithm. The classifier uses a field of power means of symmetric positive-definite matrices ([15–18]), as the previous improvement MDMF [20]. While the MDMF labels an unseen trial according to the class of the closest mean in the field, the proposed MF feeds a classifier (we have used the LDA in this study) with the squared distances to all mean matrices in the field (Figure 1). As a consequence, it can learn a complex function of these distances, while the MDMF always use the minimum distance only. The main characteristic of the proposed MF algorithm is that it is deterministic as the MDM, as long as a deterministic classifier is used.

We have benchmarked our proposition using the MOABB framework. Based on a large benchmark employing 10 MI and 10 P300 databases, MF shows very good classification performance; as compared to the state-of-the-art TS + LR Riemannian classifier, its performance is statistically lower for both MI data and P300 data, but the difference in terms of mean accuracy is little. For both MI and P300 data, the MF pipeline improves the performance significantly as compared to both the MDM and the MDMF.

The MF inherits, from its ancestor MDM, a number of desirable characteristics: it is robust, fast when combined with spatial filtering for reducing the dimension, and virtually hyperparameter-free. By 'virtually', we mean that, as compared to the MDM, it does introduce hyperparameters, namely, the number of power means and their positioning along the [−1, 1] continuum, as well as the number of filters for the ADCSP. However, these hyperparameters are not critical and reasonable a priori choices give results that are statistically indistinguishable. Further research should establish a convenient choice of these hyperparameters to be used universally.

In this article, we have also introduced an adaptive double-staged CSP filtering for MI data and shown its superiority to the standard CSP. The ADCSP greatly shortens the processing time of ensuing pipelines while also improving performance as shown in Table 3 and Figure 5 where it is compared to standard CSP. Note that the same strategy can be used for obtaining an adaptive Xdawn filtering for P300 data.

A few implementation details should be discussed. The algorithms for computing the power mean with parameter $h \in [-1, 1] \backslash 0$ and the geometric mean (limit $h \to 0$ from either side) are iterative. PyRiemann implements the algorithm in [19] for the power mean and in [49] for the geometric mean. We have used a tolerance of 1e-07 for both algorithms instead of the pyRiemann default values, which are smaller. A tolerance of 1e-07 is largely sufficient for the estimation of the means and avoids unnecessary iterations. As per the maximum number of iterations allowed, we have used 150 instead of the pyRiemann default values, which are smaller. Allowing more iterations, if needed, increases the chance to obtain a good estimate in difficult situations (e.g., low signal-to-noise ratio).

While carrying out this research, we realized that the initialization of the algorithm for power means estimation [49] was wrong in the pyRiemann implementation. This caused a slowdown in the estimations. We have corrected the bug and submitted a correction to pyRiemann. Notice that, with the correct initialization this algorithm is more robust and faster in general as compared to the one computing the geometric mean [49].

The definition of the means field itself can be optimized by adding other means that do not belong to the family of power means. Natural candidates would be, for instance, the log-Euclidean mean or $\alpha$-divergence means, to which the Bhattacharyya mean belongs [49]. Further advancements in mathematics, specifically in transposing families of means—defined for scalars—to the SPD manifold, will allow for more powerful definitions of mean fields.

A strategy for further improving the MF could be to define regions of the manifold in which means are applied. For instance, the training data may be clustered in the manifold, separately for each class, and the power means can be computed within each cluster. This way, one may compute power means for each cluster and improve the granularity of the distances submitted to the classifier.

The RPME we have adopted does improve the performance, more so for P300 than for MI data. This suggests that the P300 data included in MOABB needs to be curated. In general, open-access databases are curated and sessions with outliers are excluded. When dealing with real-world data, i.e., BCI usage out of the lab, the RPME strategy is expected to enhance the robustness of the pipeline and significantly improve performance.

## 5. Conclusions

The main contributions of this work are as follows.

First, we have proposed a fully deterministic Riemannian classifier that operates directly on the manifold of positive-definite matrices—that is, without mapping the data into the tangent space—and adopts a simple classifier, the LDA, applied to distances computed with respect to a set of power means estimated from the training data. The proposed MF classifier preserves the conceptual simplicity of the standard MDM algorithm while significantly improving performance, at the cost of only a modest increase in computational complexity.

Second, we introduced processing improvements that enhances the MF algorithm. The ADCSP improves significantly the classification performance as compared to the standard CSP. The RPME improves the robustness of the estimated means.

Our extensive analysis shows that algorithms operating only on the manifold can compete with, and potentially rival, state-of-the-art algorithms operating on tangent space that use complex classifiers. Ultimately, when a fully-deterministic and fast classifier acting on the manifold is sought, the proposed MF algorithm is, to date, the better choice.

Our code is published at: https://github.com/toncho11/The_Riemannian_Means_Field_classifier_for_EEG_Based_BCI_Data.

**Author Contributions:** Conceptualization, Anton Andreev and Marco Congedo; Methodology, Anton Andreev, Gregoire Cattan and Marco Congedo; Software, Anton Andreev and Gregoire Cattan; Writing – original draft, Anton Andreev; Writing – review & editing, Anton Andreev, Gregoire Cattan and Marco Congedo; Supervision, Marco Congedo. All authors have read and agreed to the published version of the manuscript.

**Funding:** This research received no external funding.

**Institutional Review Board Statement:** Not applicable

**Informed Consent Statement:** Not applicable

**Data Availability Statement:** The data presented in this study are contained within the article.

**Conflicts of Interest:** The authors declare no conflicts of interest.

# References

1. Chevallier, S.; Carrara, I.; Aristimunha, B.; Guetschel, P.; Sedlar, S.; Lopes, B.; Velut, S.; Khazem, S.; Moreau, T. The largest EEG-based BCI reproducibility study for open science: The MOABB benchmark. *arXiv* **2024**, arXiv:2404.15319. https://doi.org/10.48550/arXiv.2404.15319
2. Lotte, F.; Bougrain, L.; Cichocki, A.; Clerc, M.; Congedo, M.; Rakotomamonjy, A.; Yger, F. A review of classification algorithms for EEG-based brain–computer interfaces: A 10 year update. *J. Neural Eng.* **2018**, *15*, 031005. https://doi.org/10.1088/1741-2552/aab2f2
3. Levi-Civita, T. *Lezioni Di Calcolo Differenziale Assoluto*; Nicola Zanichelli Editore: Bologna, Italy, 1925.
4. Bhatia, R. *Positive Definite Matrices*; Princeton University Press: Princeton, NJ, USA, 2015.
5. Barachant, A.; Bonnet, S.; Congedo, M.; Jutten, C. Multiclass brain-computer interface classification by Riemannian geometry. *IEEE Trans. Bio-Med. Eng.* **2012**, *59*, 920–928. https://doi.org/10.1109/TBME.2011.2172210
6. Congedo, M. EEG Source Analysis. Habilitation Thesis, Université de Grenoble, Saint-Martin-d'Hères, France, 2013. Available online: https://tel.archives-ouvertes.fr/tel-00880483 (accessed on 1 February 2025).
7. Carrara, I. Advanced Methods for BCI-EEG Processing for Improved Classification Performance and Reproducibility. Phdthesis, Université Côte d'Azur, 2024. Available online: https://inria.hal.science/tel-04797267 (accessed on 12 February 2025).
8. Congedo, M.; Barachant, A.; Bhatia, R. Riemannian geometry for EEG-based brain-computer interfaces; a primer and a review. *Brain-Comput. Interfaces* **2017**, *4*, 155–174. https://doi.org/10.1080/2326263X.2017.1297192
9. Yger, F.; Berar, M.; Lotte, F. Riemannian Approaches in Brain-Computer Interfaces: A Review. *IEEE Trans. Neural Syst. Rehabil. Eng.* **2017**, *25*, 1753–1762. https://doi.org/10.1109/TNSRE.2016.2627016
10. Korczowski, L.; Congedo, M.; Jutten, C. Single-trial classification of multi-user P300-based brain-computer interface using riemannian geometry. In Proceedings of the 2015 37th Annual International Conference of the IEEE Engineering in Medicine and Biology Society (EMBC), Milan, Italy, 25–29 August 2015; pp. 1769–1772.
11. Mayaud, L.; Cabanilles, S.; Van Langhenhove, A.; Congedo, M.; Barachant, A.; Pouplin, S.; Filipe, S.; Pétégnief, L.; Rochecouste, O.; Azabou, E.; et al. Brain-computer interface for the communication of acute patients: A feasibility study and a randomized controlled trial comparing performance with healthy participants and a traditional assistive device. *Brain Comput. Interfaces* **2016**, *3*, 197–215. https://doi.org/10.1080/2326263X.2016.1254403
12. Korczowski, L.; Ostaschenko, E.; Andreev, A.; Cattan, G.; Rodrigues, P.L.C.; Gautheret, V.; Congedo, M. Brain Invaders Calibration-less P300-Based BCI Using Dry EEG Electrodes Dataset (bi2014a); GIPSA-lab, Research Report, juillet; 2019. Available online: https://hal.science/hal-02171575 (accessed on 1 February 2025).
13. Andreev, A.; Barachant, A.; Lotte, F.; Congedo, M. *Recreational Applications of OpenViBE: Brain Invaders and Use-the-Force*; chap. 14; John Wiley, Sons: Hoboken, NJ, USA, 2016. Available online: https://hal.archives-ouvertes.fr/hal-01366873/document (accessed on 1 February 2025).
14. Moakher, M. A Differential Geometric Approach to the Geometric Mean of Symmetric Positive-Definite Matrices. *SIAM J. Matrix Anal. Appl.* **2005**, *26*, 735–747. https://doi.org/10.1137/S0895479803436937
15. Lim, Y.; Pálfia, M. Matrix power means and the Karcher mean. *J. Funct. Anal.* **2012**, *262*, 1498–1514. https://doi.org/10.1016/j.jfa.2011.11.012
16. Pálfia, M. Operator means of probability measures and generalized Karcher equations. *arXiv* **2016**, arXiv:1601.06777. https://doi.org/10.48550/arXiv.1601.06777
17. Lawson, J.; Lim, Y. Weighted means and Karcher equations of positive operators. *Proc. Natl. Acad. Sci. USA* **2014**, *110*, 15626–15632. https://doi.org/10.1073/pnas.1313640110
18. Lawson, J.; Lim, Y. Karcher means and Karcher equations of positive definite operators. *Trans. Amer. Math. Soc. Ser. B* **2014**, *1*, 1–22. https://doi.org/10.1090/S2330-0000-2014-00003-4
19. Congedo, M.; Barachant, A.; Bhatia, R. Fixed Point Algorithms for Estimating Power Means of Positive Definite Matrices. *IEEE Trans. Signal Process.* **2017**, *65*, 2211–2220. https://doi.org/10.1109/TSP.2017.2649483
20. Congedo, M.; Rodrigues, P.L.C.; Jutten, C. The Riemannian Minimum Distance to Means Field Classifier. In Proceedings of the BCI 2019—8th International Brain-Computer Interface Conference, Graz, Austria, 16–20 September 2019. https://doi.org/10.3217/978-3-85125-682-6-02

21. Bruno, A.; Igor, C.; Pierre, G.; Sara, S.; Pedro, R.; Jan, S.; Divyesh, N.; Erik, B.; Barthelemy, Q.; Tibor, S.R.; et al. *Mother of All BCI Benchmarks*, version v1.0.0; 23 October 2023; Zenodo: Genève, Switzerland. https://doi.org/10.5281/zenodo.10034224
22. Jayaram, V.; Barachant, A. MOABB: Trustworthy algorithm benchmarking for BCIs. *J. Neural Eng.* **2018**, *15*, 066011. https://doi.org/10.1088/1741-2552/aadea0
23. Barachant, A.; Barthélemy, Q.; King, J.-R.; Gramfort, A.; Chevallier, S.; Rodrigues, P.L.C.; Olivetti, E.; Goncharenko, V.; Berg, G.W.v.; Reguig, G.; et al. pyRiemann, version v0.8; Zenodo: Genève, Switzerland, 2024. https://doi.org/10.5281/zenodo.593816
24. Blankertz, B.; Tomioka, R.; Lemm, S.; Kawanabe, M.; Muller, K. Optimizing Spatial filters for Robust EEG Single-Trial Analysis. IEEE Signal Process. Mag. 2008, 25, 41–56. https://ieeexplore.ieee.org/document/4408441
25. Rivet, B.; Souloumiac, A.; Attina, V.; Gibert, G. Xdawn algorithm to enhance evoked potentials: Application to brain-computer interface. IEEE Trans. Biomed. Eng. 2009, 56, 2035–2043. https://ieeexplore.ieee.org/document/4760273
26. Pham, T. Joint Approximate Diagonalization of Positive Definite Hermitian Matrices. *SIAM J. Matrix Anal. Appl.* **2001**, *22*, 1136–1152. https://doi.org/10.1137/S089547980035689X
27. Barthélemy, Q.; Mayaud, L.; Ojeda, D.; Congedo, M. The Riemannian Potato Field: A Tool for Online Signal Quality Index of EEG. *IEEE Trans. Neural Syst. Rehabil. Eng.* **2019**, *27*, 244. https://doi.org/10.1109/TNSRE.2019.2893113
28. Tangermann, M.; Müller, K.R.; Aertsen, A.; Birbaumer, N.; Braun, C.; Brunner, C.; Leeb, R.; Mehring, C.; Miller, K.J.; Müller-Putz, G.R.; et al. Review of the BCI Competition IV. *Front. Neurosci.* **2012**, *6*, 55. https://doi.org/10.3389/fnins.2012.00055
29. Leeb, R.; Lee, F.; Keinrath, C.; Scherer, R.; Bischof, H.; Pfurtscheller, G. Brain-computer communication: Motivation, aim, and impact of exploring a virtual apartment. *IEEE Trans. Neural. Syst. Rehabil. Eng*. **2007**, *15*, 473–482. https://doi.org/10.1109/TNSRE.2007.906956
30. Cho, H.; Ahn, M.; Ahn, S.; Kwon, M.; Jun, S.C. EEG datasets for motor imagery brain-computer interface. *Gigascience* **2017**, *6*, gix034. https://doi.org/10.1093/gigascience/gix034
31. Grosse-Wentrup, M.; Liefhold, C.; Gramann, K.; Buss, M. Beamforming in noninvasive brain-computer interfaces. *IEEE Trans. Biomed. Eng.* **2009**, *56*, 1209–1219. https://doi.org/10.1109/TBME.2008.2009768
32. Lee, M.H.; Kwon, O.Y.; Kim, Y.J.; Kim, H.K.; Lee, Y.E.; Williamson, J.; Fazli, S.; Lee, S.W. EEG dataset and OpenBMI toolbox for three BCI paradigms: An investigation into BCI illiteracy. *Gigascience* **2019**, *8*, giz002. https://doi.org/10.1093/gigascience/giz002
33. Schalk, G.; McFarland, D.J.; Hinterberger, T.; Birbaumer, N.; Wolpaw, J.R. BCI2000: A general-purpose brain-computer interface (BCI) system. *IEEE Trans. Biomed. Eng.* **2004**, *51*, 1034–1043. https://doi.org/10.1109/TBME.2004.827072
34. Schirrmeister, R.T.; Springenberg, J.T.; Fiederer, L.D.J.; Glasstetter, M.; Eggensperger, K.; Tangermann, M.; Hutter, F.; Burgard, W.; Ball, T. Deep learning with convolutional neural networks for EEG decoding and visualization. *Hum. Brain Mapp.* **2017**, *38*, 5391–5420. https://doi.org/10.1002/hbm.23730
35. Shin, J.; von Lühmann, A.; Blankertz, B.; Kim, D.W.; Jeong, J.; Hwang, H.J.; Müller, K.R. Open Access Dataset for EEG+NIRS Single-Trial Classification. *IEEE Trans. Neural. Syst. Rehabil. Eng.* **2017**, *25*, 1735–1745. https://doi.org/10.1109/TNSRE.2016.2628057
36. Yi, W.; Qiu, S.; Wang, K.; Qi, H.; Zhang, L.; Zhou, P.; He, F.; Ming, D. Evaluation of EEG oscillatory patterns and cognitive process during simple and compound limb motor imagery. *PLoS ONE* **2014**, *9*, e114853. https://doi.org/10.1371/journal.pone.0114853
37. Zhou, B.; Wu, X.; Lv, Z.; Zhang, L.; Guo, X. A Fully Automated Trial Selection Method for Optimization of Motor Imagery Based Brain-Computer Interface. *PLoS ONE* **2016**, *11*, e0162657. https://doi.org/10.1371/journal.pone.0162657
38. Riccio, A.; Simione, L.; Schettini, F.; Pizzimenti, A.; Inghilleri, M.; Belardinelli, M.O.; Mattia, D.; Cincotti, F. Attention and P300-based BCI performance in people with amyotrophic lateral sclerosis. *Front. Hum. Neurosci.* **2013**, *7*, 732. https://doi.org/10.3389/fnhum.2013.00732
39. Aricò, P.; Aloise, F.; Schettini, F.; Salinari, S.; Mattia, D.; Cincotti, F. Influence of P300 latency jitter on event related potential-based brain-computer interface performance. *J. Neural Eng*. **2014**, *11*, 035008. https://doi.org/10.1088/1741-2560/11/3/035008
40. Guger, C.; Daban, S.; Sellers, E.; Holzner, C.; Krausz, G.; Carabalona, R.; Gramatica, F.; Edlinger, G. How many people are able to control a P300-based brain-computer interface (BCI)? *Neurosci. Lett*. **2009**, *462*, 94–98. https://doi.org/10.1016/j.neulet.2009.06.045
41. Van Veen, G.F.P.; Barachant, A.; Andreev, A.; Cattan, G.; Rodrigues, P.L.C.; Congedo, M. Building Brain Invaders: EEG Data of An Experimental Validation. GIPSA-Lab, Research Report 1, mai 2019. Available online: https://hal.archives-ouvertes.fr/hal-02126068 (accessed on 1 February 2025).

42. Vaineau, E.; Barachant, A.; Andreev, A.; Rodrigues, P.C.; Cattan, G.; Congedo, M. Brain Invaders Adaptive Versus Non-Adaptive P300 Brain-Computer Interface Dataset. GIPSA-LAB, Research Report 1, avril 2019. Available online: https://hal.archives-ouvertes.fr/hal-02103098 (accessed on 2 July 2019).
43. Korczowski, L.; Ostaschenko, E.; Andreev, A.; Cattan, G.; Rodrigues, P.L.C.; Gautheret, V.; Congedo, M. Brain Invaders Solo Versus Collaboration: Multi-User P300-Based Brain-Computer Interface Dataset (bi2014b); GIPSA-lab, Research Report, juillet; 2019. Available online: https://hal.science/hal-02173958/ (accessed on 1 February 2025)
44. Korczowski, L.; Cederhout, M.; Andreev, A.; Cattan, G.; Rodrigues, P.L.C.; Gautheret, V.; Congedo, M. Brain Invaders Calibrationless P300-Based BCI with Modulation of Flash Duration Dataset (bi2015a); GIPSA-lab, Research Report, juillet; 2019. Available online: https://hal.science/hal-02172347 (accessed on 1 February 2025).
45. Korczowski, L.; Cederhout, M.; Andreev, A.; Cattan, G.; Rodrigues, P.L.; Gautheret, V.; Congedo, M. Brain Invaders Cooperative Versus Competitive: Multi-User P300-based Brain-Computer Interface Dataset (bi2015b); GIPSA-lab, Research Report, juillet; 2019. Available online: https://hal.science/hal-02173913v1 (accessed on 1 February 2025).
46. Edgington, E.S. *Randomization Tests*, 3rd ed.; Marcel Dekker: New York, NY, USA, 1995.
47. Huson, L.W. Multivariate Permutation Tests: With Applications in Biostatistics. *J. R. Stat. Soc. Ser. D Stat.* **2003**, *52*, 247. https://doi.org/10.1111/1467-9884.t01-6-00356
48. Congedo, M.; Afsari, B.; Barachant, A.; Moakher, M. Approximate Joint Diagonalization and Geometric Mean of Symmetric Positive Definite Matrices. *PLoS ONE* **2015**, *10*, e0121423. https://doi.org/10.1371/journal.pone.0121423